\documentclass[longbibliography]{iopconfser}

\usepackage{graphicx}
\usepackage{subfigure}
\usepackage{color}

\usepackage{ulem}

\begin{document}

\title{Preservation of the Direct Photon and Neutral Meson Analysis in the PHENIX Experiment at RHIC}

\author{Gabor David$^{1}$, Maxim Potekhin$^{2}$ and Dmitri Smirnov$^{2}$}

\affil{$^1$Department of Physics and Astronomy, Stony Brook University, Stony Brook, NY, USA  \\
$^2$Department of Physics, Brookhaven National Laboratory, Upton, NY, USA}


\email{david@bnl.gov}

\begin{abstract}
The PHENIX Collaboration has actively pursued a Data and Analysis Preservation program since 2019, the first such dedicated effort at RHIC. A particularly challenging aspect of this endeavor is preservation of complex physics analyses, selected for their scientific importance and the value of the specific techniques developed as a part of the research. For this, we have chosen one of the most impactful PHENIX results, the joint study of direct photons and neutral pions in high-energy d+Au collisions. To ensure reproducibility of this analysis going forward, we partitioned it into self-contained tasks and used a combination of containerization techniques, code management, and robust documentation. We then leveraged REANA (the platform for reproducible analysis developed at CERN) to run the required software. We present our experience based on this example, and outline our future plans for analysis preservation.
\end{abstract}

\section{Introduction}


The PHENIX experiment~\cite{PHENIX:2003nhg} was one of the two major experiments at the Relativistic
Heavy Ion Collider (RHIC) at Brookhaven National Laboratory (BNL). It was built in the 1990s and
started taking data in 2000, playing the leading role in the discovery of the Quark-Gluon
Plasma (QGP) and studying its properties~\cite{PHENIX:2001hpc,PHENIX:2003qra,PHENIX:2004vcz}. 
The experiment continued to operate through 2016 and recorded a total of 24PB of raw data,
covering a large variety of ions and collision energies, made possible by the versatility of RHIC.

While PHENIX has already published more than 220 physics papers, a few important analyses are still ongoing even now, 
and there is a possibility of new analyses that may start in the future, triggered by new physics insights or
emergence of new analysis methods and tools.  This is not uncommon 
-- see for instance the recent re-analysis of HERA H1 data taken 2006-2007~\cite{Mikuni:2024hzj}. However, in the era of rapidly changing hardware and software environments 
and fading institutional memory, 
ensuring that the data 
can be meaningfully 
analyzed decades from now is a very complex task. This work area is usually referred to as Data and Analysis Preservation (DAP).
Ideally, DAP should allow a completely new analysis of the data (new observable or method), but reaching a reduced goal,
namely making an existing and published analysis fully reproducible at any later time is already an important achievement.  

\begin{figure}[htbp]
    \centering
    \includegraphics[width=0.50\linewidth]{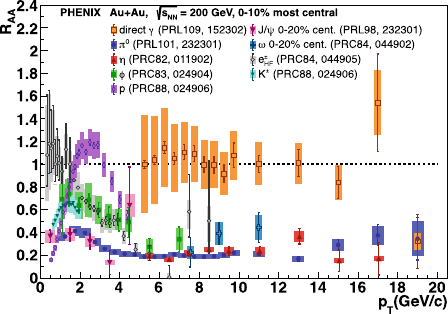}
    \caption{Nuclear modification factor in 200\,GeV Au+Au collisions for various hadrons and direct photons.  }
\label{fig:T-shirt}
\end{figure}

In this paper we report on the successful preservation of the high transverse momentum direct
photon and $\pi^0$ analysis~\cite{daupaper:2023} of the 200\,GeV $d$+Au collision data that was taken in 2016.
As seen in the iconic Fig.\ref{fig:T-shirt} measurement of the nuclear modification factor, particularly
for neutral mesons and direct photons is one the supreme capabilities of PHENIX.
Also, observation of the large hadron suppression and non-suppression of photons was one of the
groundbreaking results at RHIC, decisive in establishing the formation of the QGP in large ion collisions.
Apart of being highly interesting in its own right, we have chosen the $d$+Au analysis as our first DAP target, because this is the first time $\pi^0$ and $\gamma$ have been analyzed and published in the same workflow, intertwined, and with a method that substantially reduces previous systematic uncertainties.  Also, the analysis has important implications to the determination of collision centrality in small system collisions.


\section{Summary of the Data and Analysis Preservation effort in the PHENIX experiment}

Data and Analysis Preservation (DAP) became a focus area for the Collaboration in 2019. At its core it has three principal components~\cite{Potekhin:2023zkb}
\begin{itemize}
    \item Data preservation, including raw/processed data, databases, metadata
    \item Management and preservation of necessary core and analysis-specific software and its configuration
    \item Knowledge management: capturing and making available all relevant information even after the original analyzers are not available and "institutional memory" fades
\end{itemize}

\subsection{Data preservation}

PHENIX~\cite{phenixpubweb} relies on its host (Brookhaven National Laboratory) for archiving of its raw data and a subset of the derived data products~\cite{Hollowell:2017jjd}. A special area within the disk mass storage has been allocated for keeping intermediate data crucial for Analysis Preservation, including the analysis of direct photons in d+Au collisions. Conditions and calibrations data are stored in a database, and the issue of preserving these data will be addressed at a later time. The publication-level data such as data points used in tables in plots is reliably preserved by PHENIX Collaboration, by leveraging the HEPData Portal at CERN. A vast majority of all published PHENIX results are now preserved in this manner.

\subsection{Software preservation}



The PHENIX software stack, the result of over 25 years of continuous development, is inherently complex and includes a number of legacy components. Preserving this software presents significant challenges, particularly in maintaining compatibility amid ongoing component upgrades and evolving operating systems. To address these challenges, the PHENIX Collaboration has adopted containerization technology, specifically Docker, which provides an efficient solution for preserving the software environment. Docker enables the encapsulation of the entire software stack, ensuring both its continued operation and portability across different systems.

In addition to containerization, we have prioritized reproducibility in our computational workflows for physics analysis. To achieve this, we integrated the REANA platform~\cite{Simko:2018zzz}, which facilitates the orchestration of complex workflows within a containerized environment. REANA not only ensures that our analyses are reproducible but also that they can be reliably executed in the future, irrespective of changes in the underlying infrastructure.

This combined approach of Docker-based containerization and the REANA platform has proven effective in overcoming the challenges of software preservation, ensuring that the PHENIX software stack remains functional and accessible for future analyses.

\subsection{Knowledge management}

Even a perfectly preserved software can’t be useful without accompanying “know-how” for specific applications and related configuration details. This is being addressed via the PHENIX Knowledge Management effort. PHENIX has successfully migrated a large number – more than 620 –  of its research documents (including PhD theses, technical write-ups, conference presentations etc) to a modern digital repository — the Zenodo platform at CERN~\cite{phenixzenodo}. The previously fragmented Web materials have been curated and migrated to a new website, created with the goals of simplicity, security, performance and ease of long-term management~\cite{phenixpubweb}. All these measures combined allowed us to create a functioning foundation for Analysis Preservation.

\section{The Analysis of $\pi^0$ and direct $\gamma$ in d+Au collisions}

\subsection{The physics issues addressed in the analysis}
The complete description of the context, significance, issues and physics consequences drawn from the analysis of $d$+Au data are beyond the scope of this paper and are described in~\cite{daupaper:2023}, as well as in a dedicated write-up for this analysis~\cite{ana_zenodo}. Some of the key results in heavy-ion experiments such as PHENIX
are the nuclear modification factors (essentially the qualitative difference between particle production in heavy ion vs. proton-proton collisions) which are studied as a function of centrality  (the amount of overlap between the colliding ions). Conclusions are based in a large part on the measured invariant yields (spectra) of $\pi^0$ and direct $\gamma$ as a function of centrality.

The steps of the analysis leading to those invariant 
$\pi^0$ and direct $\gamma$ yields are shown in Fig.~\ref{fig:workflow} and preserved in our REANA implementation. This workflow is designed such that can be run on a batch system and has been demonstrated to reproduce the fundamental quantities being studied, such the spectra in~\cite{daupaper:2023}, thus satisfying the first and foremost goal of DAP: reproducibility of a published result.  This is illustrated in Fig.~\ref{fig:reanaout}.  While the analysis as a whole is preserved, some of its parameters, cuts and conditions in the individual macros can be changed, making somewhat different analyses of the same data possible, so it can be viewed as a meaningful step toward the more ambitious, extended DAP goal of new analyses of preserved old data.  The entire workflow and all relevant code, along with the necessary configuration files is preserved is a version control system accessible to users with BNL SDCC credentials here~\cite{ana_gitea}, with the necessary input files stored in~\cite{ana_inputfiles}. The foundational software stack used in PHENIX has been preserved in a Docker image,  ensuring portability.

\subsection{The REANA implementation}

\begin{figure}[htbp]
    \centering
    \includegraphics[width=0.8\linewidth]{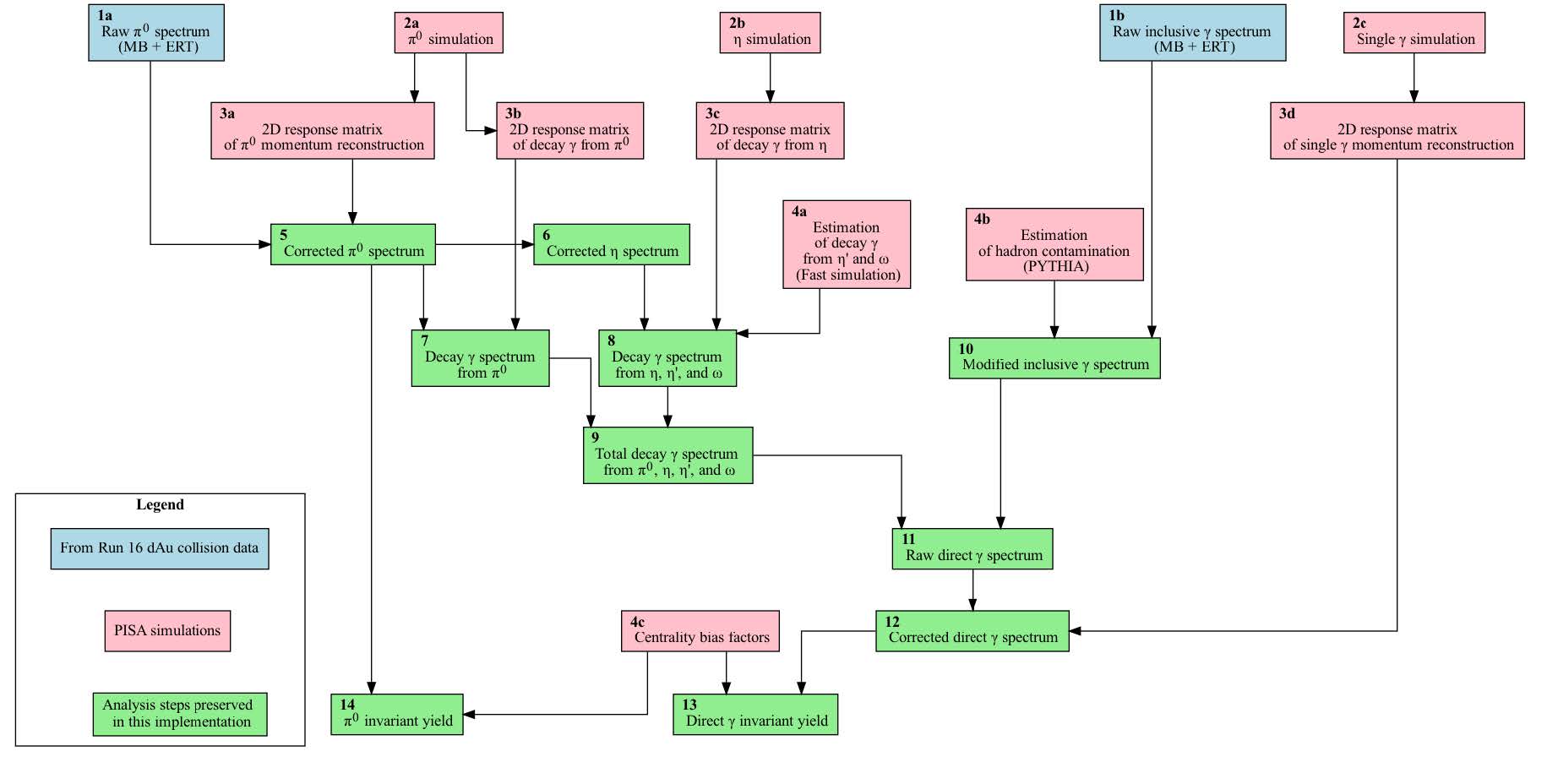}
    \caption{Workflow of the $\pi^0$ and direct $\gamma$ analyses in PHENIX, as preserved. See the text for details.}
\label{fig:workflow}
\end{figure}

The REANA implementation of our analysis workflow, as illustrated in Fig.~\ref{fig:workflow}, is designed to encapsulate and automate the complex sequence of steps required for data analysis in PHENIX.

In this workflow, blue boxes represent operations performed on a condensed output derived from a comprehensive scan of all available data, stored as a series of multidimensional histograms. These condensed outputs, essential for various stages of the analysis, are stored within the containerized environment. Pink boxes indicate operations involving simulated particles embedded in real events, generating what are known as "response matrices," which are 2D histograms that capture the detector's response to different physical scenarios. These matrices are also preserved within the container, ensuring that all necessary inputs for analysis are readily accessible. Green boxes correspond to the actual analysis steps, which involve executing the preserved ROOT macros. Analysts can edit these macros to modify parameters, apply different cuts, or adjust conditions according to their specific analysis objectives. Each step in the workflow is accompanied by a verbal description to guide users through the process.

\begin{figure}
    \subfigure{\includegraphics[width=0.48\textwidth]{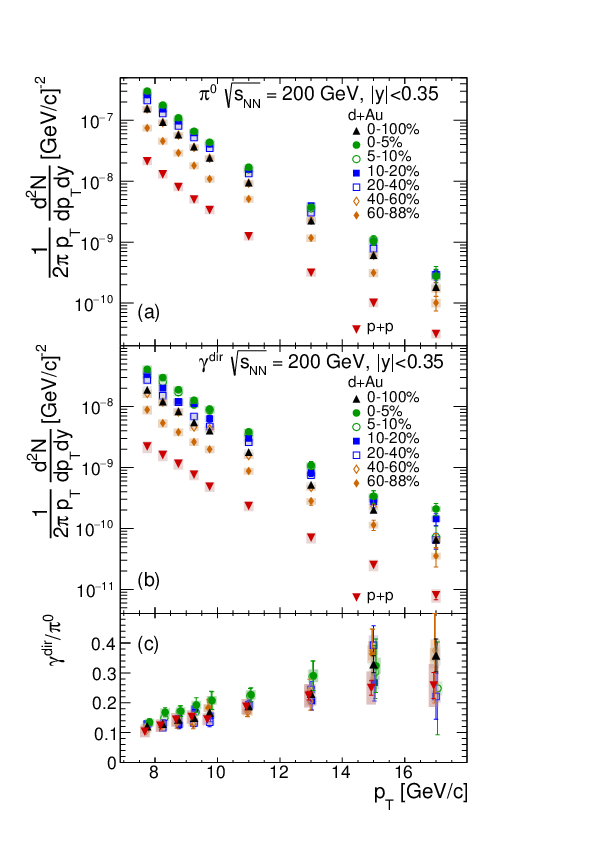}}%
    \subfigure{\raisebox{5mm}{\includegraphics[width=0.41\textwidth]{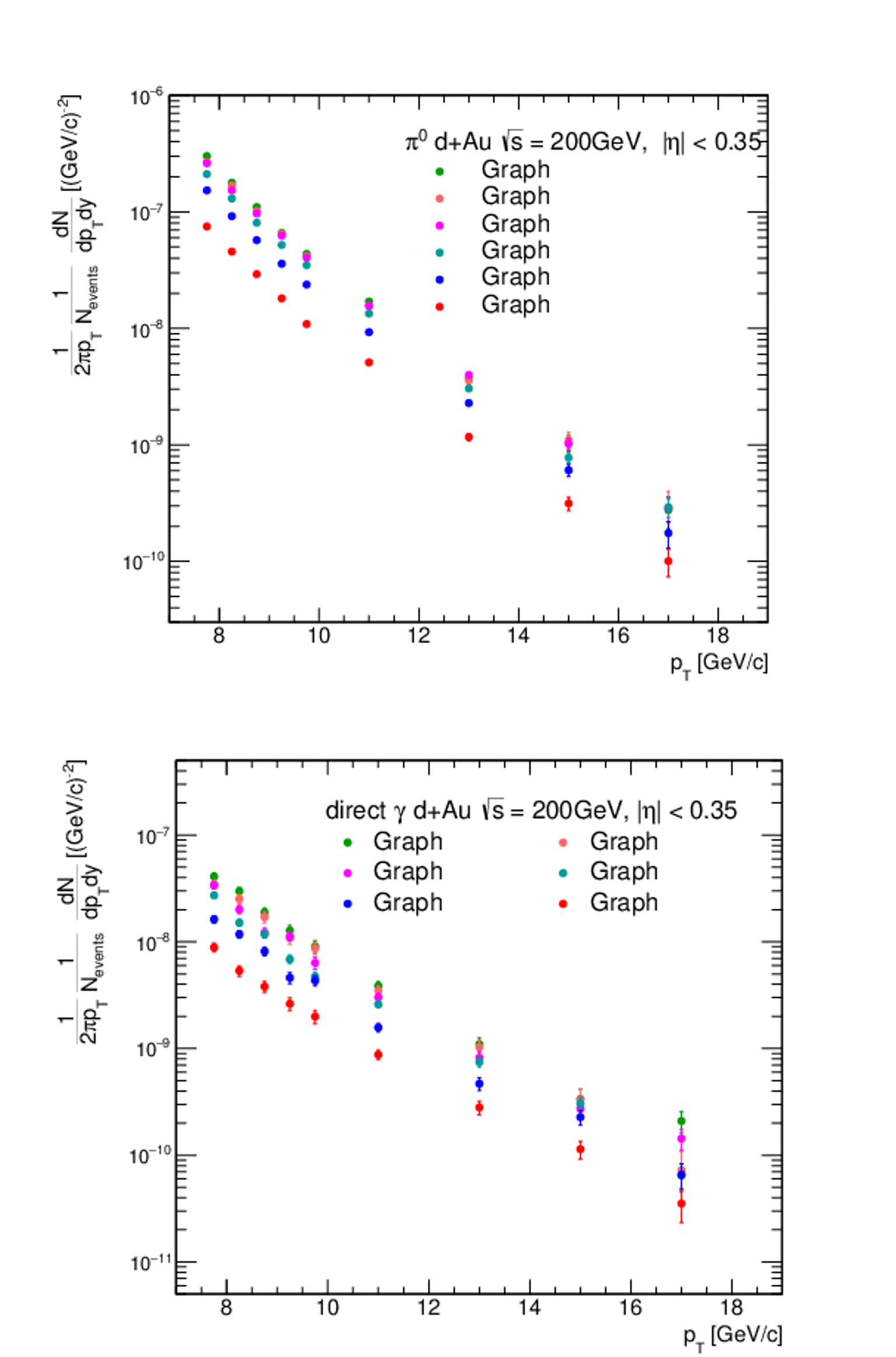}}}%
    \caption{Left: final results of $\pi^0$ and direct $\gamma$ production as a function of centrality in 200\,GeV $d$+Au collisions~\cite{daupaper:2023}.  also shown are the corresponding spectra for 200\,GeV $p$+$p$ collisions, taken from a different analysis, and a derived quantity, the point-by-point ratio of the $\pi^0$ and $\gamma$ spectra (bottom panel).  Right: $\pi^0$ (top) and $\gamma$ spectra as reconstructed in the REANA implementation shown in Fig.~\ref{fig:workflow}. }
    \label{fig:reanaout}
\end{figure}

To validate the robustness of our preservation approach, the containerized analysis was tested by a computer-literate individual who is not a PHENIX collaborator. This individual successfully executed the entire analysis chain, with the results shown in Fig.~\ref{fig:reanaout}. The left panel displays the $\pi^0$ and direct $\gamma$ spectra (and their ratio) as originally published, while the right panel shows the same spectra generated from the preserved analysis chain. This successful replication by an external user underscores the effectiveness of the REANA implementation in preserving not just the data, but also the full analytical capability for future research.


\subsection{Lessons learned}
The PHENIX Collaboration has been largely successful in its Data and Analysis Preservation
program (DAP),
however this required a substantial amount of effort, against the background of limited
personnel available for work in this area. Adding to the challenge, the PHENIX hardware and
software design started in early 1990s, and for obvious reasons the software environment
underwent major changes since then. Over the years, the software stack became very large,
with complex code dependencies and in some cases redundancies. In view of this, we conclude
that the single most important lesson from our efforts is that a well-designed, long-term DAP
component of the experiment needs to be created by professionals from the very beginning of
the experiment and the required standards meticulously followed during its entire lifetime,
even if it appears costly or counterproductive at any particular moment. We recognize that
a robust analysis preservation effort may be considered a hurdle on the path to quickly
obtaining new and exciting publishable results. For that reason, another important lesson
is that clear, stable guidance is needed from the very beginning, on how to balance there
contradictory requirements. One other dimension of the work we are doing is balancing
the legitimate requirements of openness (since the data, the code and the analysis
techniques need to be eventually made public due to the policies of the funding
agencies), against the also legitimate requirements of data protection and
cybersecurity. Reconciling all of this adds to the already significant challenge,
and once again, clear guidance and sufficient funding are instrumental in
achieving the worthy goals of the Data and Analysis Preservation.

\section{Summary}
The PHENIX experiment stopped taking data in 2016, having accumulated a large amount of
data, much of it with a unique physics reach. Analyses of these data are ongoing and even
new analyses can be expected to take place in the future, requiring a long term Data and Analysis Preservation effort. Such effort started in PHENIX in 2019 with limited manpower. Substantial progress
has been achieved in the area of Knowledge Management, leveraging a web technology suited for long-term maintenance and a modern digital repository.
Data tables for almost all PHENIX publications are now available on the HEPData portal~\cite{hepdata}.
The PHENIX software environment has been preserved with the use of containerization technology.
Recently the analysis chain of a high-impact PHENIX paper has been implemented in REANA in the containerized format. The next project is REANA implementation of another signature PHENIX analysis, charmonium production via the
$J/\Psi\rightarrow\mu\mu$ decay channel.






\bibliographystyle{IEEEtran}
\bibliography{acat2024}

\end{document}